\documentclass{article}
\usepackage{graphicx}
\begin{document}
\bibliographystyle{plain}
\vspace{1.5cm}
\begin{center}
{\LARGE \bf
Dynamic re-wiring of protein interaction:\\ The case of transactivation}
\vspace{1cm}

{\Large \it Gabriele Scheler\\ISLE\\Ventura Hall 25\\200 Panama Street\\Stanford, Ca. 94305}

\end{center}

\begin{abstract}
We are looking at local protein interaction networks from the perspective 
of directed, labeled graphs with quantitative values for monotonic changes in 
concentrations. These systems can be used to perform stability analysis for 
a stable attractor, given initial values. They can also show 
re-configuration of whole system states by dynamic insertion of 
links, given specific patterns of input.  The latter issue seems particularly 
relevant for the concept of multistability in cellular memory.
We attempt to show that this level of analysis is well-suited 
for a number of relevant biological subsystems, such as transactivation in 
cardiac myocytes or G-protein coupling to adrenergic receptors. In particular, 
we analyse the 'motif' of an "overflow gate" as a concentration-dependent 
system reconfiguration.
\end{abstract}
\vspace*{1.0cm}
\section{Introduction}

We are interested in cases, where the intracellular pathways established by 
protein interaction and small messenger molecules show a dynamic change 
('re-wiring') due to external conditions. For two distinct major pathways, the 
G-protein coupled pathway activated by ligand-bound receptor molecules and 
the MAP kinase pathway activated by tyrosine kinases, it has been shown 
that regulatory interactions arise under certain conditions,
in a process known as 'transactivation'(\cite{GschwindAetal2001,ShahCatt2004}). 
Transactivation is thus an ideal model 
case to study dynamic re-wiring of protein interactions, its conditions, and 
its functional consequences. Another related process, the antagonism of 
G-protein coupled receptors, which occurs as a submodule in transactivation,
provides another opportunity to analyse the dynamics of protein interaction.

The computational analysis of protein interaction has focused on the 
establishment of protein interaction networks and their analysis by search for 
local motifs or subgraphs in attempts to link specific subgraphs to functional 
modules \cite{MiloRetal2002}. Showing how and why functional subgraphs change 
will add a much needed dynamical perspective to the analysis of protein 
interaction networks. 
To analyse these patterns seems a useful step in order to cut down on the 
complexity of the temporal dynamics of intracellular signaling 
in a given situation.

In many ways, intracellular networks need to function as adaptive control 
systems: keeping key parameters within tightly controlled bounds in response 
to many extracellular events. However, to provide memory, they also need to 
be {\it multistable}, i.e. function within different regimes. 
Probabilistic re-wiring of subgraphs provides an adequate representation for
the dynamics of interactions to allow to analyse changing 
control pathways and their stability (cf. \cite{SchelerSchumann2004}).

\section{Transactivation of the MAP kinase pathway by G proteins:
A cascade of overflow gates}

Tyrosine kinases (such as EGFR) occur in membrane-bound position where they can be activated 
by both extracellular and intracellular events. They comprise an evolutionarily 
conserved, 'old' type of receptor with little specificity in conditions 
for activation (they are typically activated by up to six different ligands)
and effects on an important intracellular pathway, 
the MAP kinase pathway, which is involved in cellular housekeeping 
functions, such as initiating growth or apoptosis.

The large protein family of G-protein coupled receptors (GPCRs) consists of 
membrane-bound proteins which change their conformational state by being bound 
with specific ligands in extracellular concentrations and effecting the separation 
of oligomeric G proteins ($G_s$, $G_q$ and $G_i$ proteins) into $G_{\alpha}$ 
and $G_{\beta\gamma}$ components \cite{PreiningerHamm2004}.
There are usually at least 2, up to 7 or so
different GPCRs that bind to the same extracellular ligand with considerable 
specificity.
The different G proteins they interact with have, among other effects, antagonistic functions 
on downstream signaling: $G_{s}$ proteins augment the adenylyl 
cyclase, cyclic AMP and protein kinase A pathway and 
$G_{i}$ proteins suppress this pathway. 

\begin{figure}[htb]
\includegraphics[width=5cm] {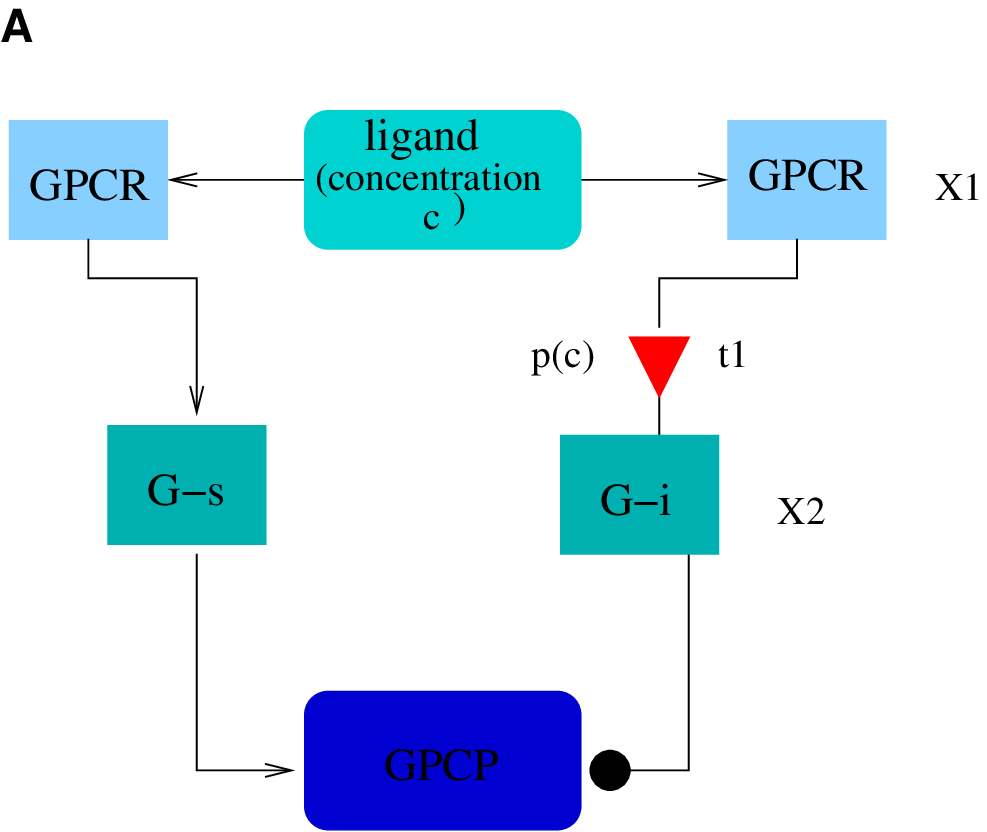}\hspace*{2.0cm}
\includegraphics[width=5cm]{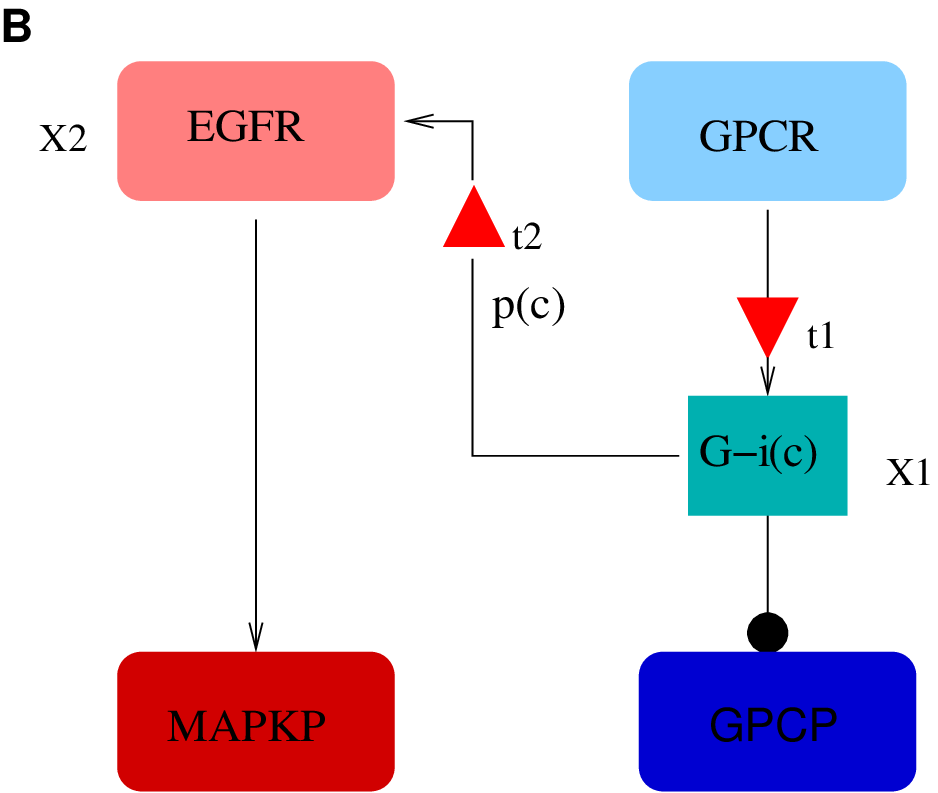}
\caption{Two examples for overflow gates (A) G-i coupled proteins become 
activated at high concentrations of extracellular ligand (B) transactivation 
of a receptor tyrosine kinase requires high levels of G-i proteins. 
Links are inserted (gated) with a concentration-dependent probability function 
$p(c)$} 
\label{figure-1}
\end{figure}

We propose that the basic function of this GPCR antagonism is a regulation of 
parametric ranges for the common substrate in downstream signaling, the 
cAMP/PKA pathway('GPCP'). 
It is remarkable that for each ligand, there exists usually both 
 an 'excitatory', $G_{s}$-coupled receptor, and an
'inhibitory' $G_{i}$ coupled receptor 
\cite{PreiningerHamm2004}.
Mostly, the Gi-coupled receptor has a lower 
affinity to the ligand, and thus becomes activated only at higher 
concentrations. Thus there is an overflow-dependent activation of an 
additional protein interaction, in this case with a suppressive effect 
on a common outcome with a pre-existing interaction 
(see Fig.~\ref{figure-1}A).

The basic principle in 'transactivation' relies on a variation of the 
same concept: 
A high level of 
$G_i$ protein activity leads to the insertion of an interactive link from 
$G_{\beta\gamma}$ proteins to the EGF receptor, resulting 
in an additional, augmenting activation of this receptor 
\cite{GschwindAetal2001,ShahCatt2004}. 
Functionally, this corresponds to the insertion of a regulatory 
interaction between the GPC and the MAPK pathways triggered by the 
concentration of a protein in one of 
the pathways (see Fig.~\ref{figure-1}B,t2).

We want to establish this basic mechanism of a 
concentration-activated protein interaction in very general terms, as a 
simple, conserved motif in the dynamics of protein interaction:
\begin{itemize}
\item[1.]
A link between $X_1$ and $X_2$ becomes established with probability $p$, 
where $p$ 
depends in a functional manner on a concentration $c$ related to $X_1$.
(concentration-dependent gate, overflow gate)
\end{itemize}
The concentration $c$ may be identical to a local concentration of 
$X_1$ as in Fig.~\ref{figure-1}B or the concentration of an extracellular 
ligand (or small molecule, such as cAMP or calcium) that determines 
the activation state (conformational change) of $X_1$ 
(as in Fig.~\ref{figure-1}A).
The probability p will in many cases be determined by a sigmoidal 
function, which establishes 
a lower threshold for the probability to increase and an upper threshold 
(saturation) for it to remain constant (U-shaped or other functions are also 
possible). A local protein 
interaction network may operate preferentially in a low regime - where the 
sigmoidal function operates mainly as a threshold - or in a higher regime - 
where the sigmoidal function operates mainly as a linear dependence.

There are potentially other motifs, which require only a probabilistic 
interpretation of concentrations: e.g. a pool concentration may interact with 
two different partners according to a fixed partitioning 
($\theta$, $1-\theta$), with the possibility of one of these connections 
acting as an informative control signal, 'telling' another pathway about 
the level of activation of another pathway (Example: 
\cite{BarronAJetal2003,
LeeKHetal2003,KaroorVetal2004}).  

Note however that the choice of a definition involving probabilities in 
determining 
protein interaction does not capture the time course of interactions 
explicitly
(see below, section \ref{temporal} for a discussion).

\begin{figure}[htb]
\includegraphics[width=6cm]{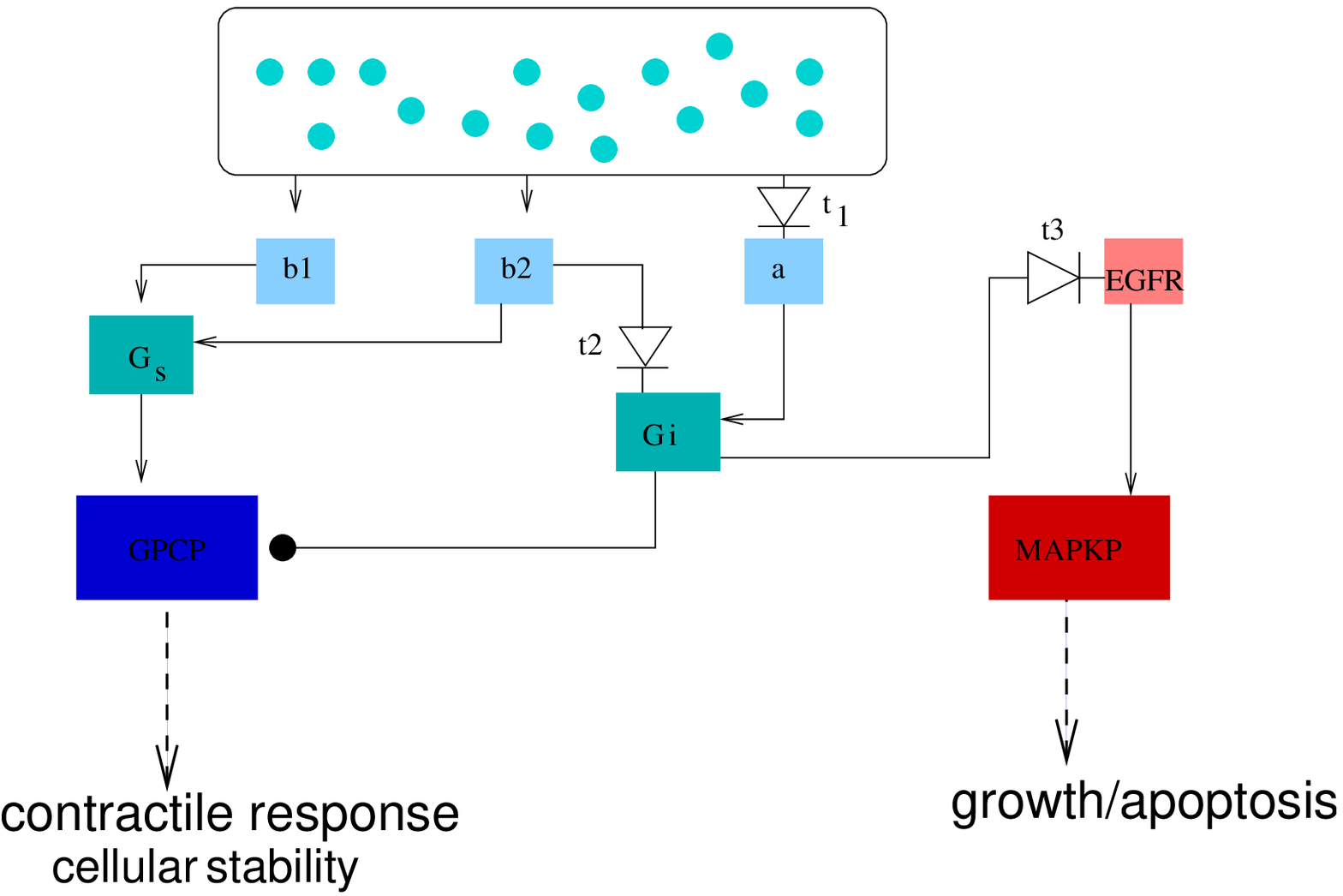}\hspace*{2.0cm}
\includegraphics[width=5cm]{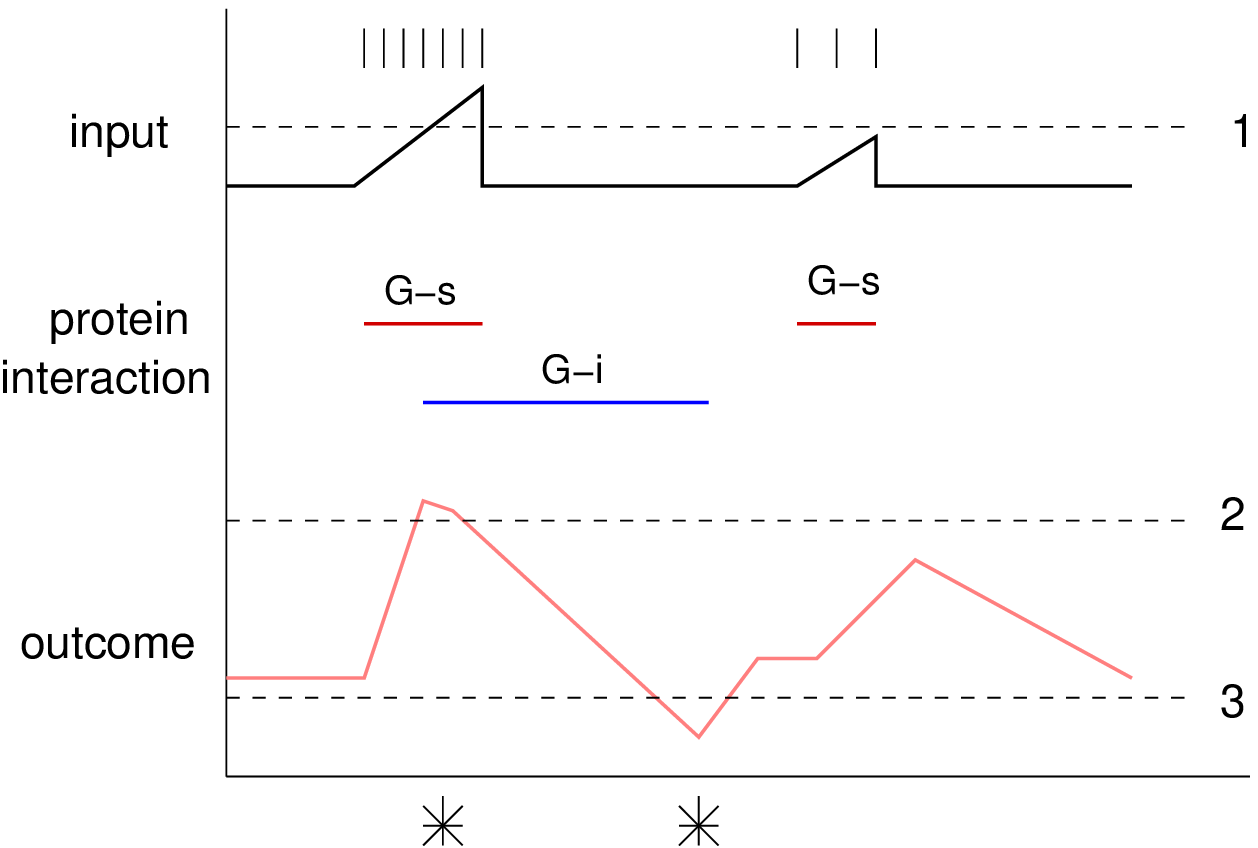}
\caption{(A) A simple graphic showing three 'overflow gates'(t1-t3). A high 
extracellular signal activates first t1 and then t2. A high G protein signal 
activates t3. Cellular outcome with overactivation of the MAPK pathway is 
cellular growth or apoptosis, otherwise stability is maintained. (B)
A quantal input, linearly accumulating, (top) interacts with proteins with a threshold (line1) for 
the 'overflow gate' G-i. The time course of the outcome (GPCP) shows 
a short period of overactivation (first *) and hysteresis (second *) relative
to the optimal parametric range (between line2-3).}
\label{figure3}
\end{figure}

Transactivation occurs in many types of cells \cite{ShahCatt2004}, and involves a number of 
additional processes \cite{ShahCatt2004,Sugden2001}. Within the context of 
cardiac myocytes, another prominent process is aptly categorized as an 
'overflow gate':
The native ligand, noradrenaline, activates both 
$\alpha$- ($G_i$-coupled) and $\beta$-($G_s$-coupled) adrenergic receptors, 
which are antagonistically related.
In addition, there are two different variants of $\beta$-adrenergic receptors, 
$\beta$-1 and $\beta$-2, where $\beta$-2 additionally activates 
$G_i$ proteins, when being continually or 
highly activated \cite{ZamahAMetal2002}. 
In cardiac myocytes, different cellular outcomes (increased contractile
responses \cite{CommunalCetal2000}, 
growth and apoptosis 
\cite{SinghKetal2001,HenaffMetal2000})
are
associated with the level of  
activation of the MAPK pathway system. We see, interestingly, that the 
overactivation associated with a harmful response is being protected by 
a system consisting of a three-degree cascade of overflow 
gates (see Fig.~\ref{figure3}A).

\section{Temporal dynamics of an 'overflow gate'}
\label{temporal}
By focusing on the {\it probability} of a protein interaction, we lose the 
temporal fine structure of concentration changes in establishing a link that 
can be derived from kinetic reaction rate models or similar dynamical 
equations.
For instance, not only are G-s coupled receptors activated by lower 
levels of ligands, 
they also desensitize more quickly, and undergo faster recycling at the 
membrane.  G-i coupled receptors, which require higher levels of ligands,
desensitize more slowly, and have slower turnover rates as well.
The fine temporal structure of this process is depicted schematically in 
Fig.~\ref{figure3}B. 
Initial G-s activation is counteracted after a delay by G-i activation 
if the signal keeps increasing. G-i activation lasts for a considerable longer 
period of time which may result in hysteresis of the response. 
Instead of G-s and G-i interacting to keep a stable range of outcome no 
matter how strong the signal, we now have time courses for the outcome 
where high values may be tolerated, if they are counteracted in time 
by an antagonistic force.
We can see that the requirements of stable control 
parameters can be temporarily lifted to create significant high or low 
signals.
The signal also has a characteristic time course 
with a rapid increase and a slow dynamics of suppression (common in a system 
of antagonistic regulation). Other patterns for time courses are, for instance,
slow rise times, which correspond to temporal buffers.
Thus the fine temporal structure can also 
be analysed with the goal of discovering most simple motifs  
with considerable generality. 
Biochemically, protein interactions can be adequately modelled and simulated 
by reaction-rate equations, such as the Michaelis-Menten formulation.
Integrating explicit rate information into a probabilistic dynamic structure 
also bridges the gap to an actual simulation model of protein interaction.

\section{Summary}

We have started from a well-described model system (transactivation in 
cardiac myocytes) in an attempt to find simple, general 
dynamic motifs, beyond the paradigm case of the 'feedback loop'. We have 
also outlined 
how this information can be included in existing 
protein interaction networks (e.g. \cite{ChinSamanta2003}), using a 
probabilistic formulation for concentration-dependent dynamic re-wirings, 
and specifically analysed the motif of an 'overflow switch' with three examples.

Actual protein interaction occurs in tight local structures
(using scaffolds and other properties of the actin cytoskeleton) which 
often give the impression of highly ordered, well-designed machinery.
This is particularly true for membrane-bound and membrane-close proteins 
(in contrast to cytosolic proteins which diffuse more freely). 
Protein networks only provide a static picture of possible interactions. 
A dynamic view, however, can be achieved on different levels. 
One goal is to describe control interactions 
and adaptive stability or multistability within the system.
We have tried to show how to identify and define dynamic motifs
which do two things: they insert a link in a subgraph and 
they specify the conditions for the link to become active.
For this we used quantities(local concentrations) and (sigmoidal or other)
probabilities of interaction.
This allows the specification of memory in the system, and goes beyond a 
static description of all possible interactions, and their characteristics. 

To fully reconstruct a complex biological 
function such as neural adaptivity, we will also need to look at the 
time-courses of changes and integrate over different time-scales. 
Thus, another goal of a dynamic network is to provide an integration with 
simulation models on the level of  kinetic rate equations. Here again, 
we may look for conserved motifs of characteristic time curves and their 
associated function.

Engineering and evolution share an approach of 
building complex systems from repetition of
simple modules \cite{SpirinMirny2003}, \cite{JordanJDetal2000}.
The real challenge lies in understanding the integrative principles 
in cellular systems, which may be significantly different from the
planned design approach characteristic of engineered systems.

\bibliography{transact}
\end{document}